\documentclass[12pt]{article}
\usepackage{epsfig}
\usepackage{amssymb}
\usepackage{amsmath}
\usepackage{bm}                  
\usepackage{cancel}
\usepackage{graphics}
\usepackage{comment} 
\newcommand{\pres}[2]{\setbox0=\hbox{$\scriptstyle #1$} \dimen0=\dp0  
              \dimen1=\ht0 \divide\dimen1 by 3
              \advance\dimen0 by \dimen1
              \hbox{\lower\dimen0 \hbox{$\scriptstyle #2\!\!$}} #1}
\newcommand{\nc}{\newcommand}
\nc{\figcap}[1]{\begin{quote}\refstepcounter{figure}
        {\bf Figure \thefigure}: {\small #1}\end{quote}}
\nc{\fig}[1]{\mbox{Fig.~\ref{#1}}}
\nc{\noi}{\noindent}
\nc{\bea}{\begin{eqnarray}}
\nc{\eea}{\end{eqnarray}}
\nc{\bean}{\begin{eqnarray*}}
\nc{\eean}{\end{eqnarray*}}
\nc{\ba}{\begin{array}}
\nc{\ea}{\end{array}}
\nc{\be}{\begin{equation}}
\nc{\ee}{\end{equation}}
\nc{\nn}{\nonumber}
\nc{\bra}[1]{\langle #1|}
\nc{\ket}[1]{|#1\rangle}
\nc{\av}[1] {\langle #1\rangle}
\nc{\vac}[1] {\langle 0| #1|0\rangle}
\nc{\amp}[2]{\langle #1|#2\rangle}
\nc{\da}{\dagger}
\nc{\pa}{\partial}
\nc{\ga}{\gamma}
\nc{\ep}{\epsilon}
\nc{\tf}{t_f}
\nc{\half}{\ensuremath{\frac{1}{2}}}
\nc{\hHH}{\hat H}
\nc{\ha}{\hat a}
\nc{\hO}{\hat O}
\nc{\hAA}{\hat A}
\nc{\hB}{\hat B}
\nc{\hG}{\hat G}
\nc{\hN}{\hat N}
\nc{\hU}{\hat U}
\nc{\hx}{\hat{x}}
\nc{\hp}{\hat{p}}
\nc{\hpsi}{\hat \psi}
\nc{\hphi}{\hat \phi}
\nc{\hpi}{\hat \pi}
\nc{\hpd}{\hat \psi ^\dagger}
\nc{\hE}{\hat E}
\nc{\hb}{\hat b}
\nc{\hc}{\hat c}
\nc{\hjo}{\hat j _0}
\nc{\hrho}{\hat \rho}
\nc{\leave}{\! \! \! \! \! / \, \,}
\nc{\intl}[1]{\int d\! #1 \,} 
\nc{\intll}[3]{\int _#1^#2 d\! #3 \,} 
\nc{\lm}{\lim _{y \rightarrow x}}
\nc{\scd}{\partial ^2 _{A_T}}
\nc{\fd}[1]{\frac{\delta }{\delta #1}} 
\nc{\pad}[1]{\frac{\partial}{\partial #1}} 
\nc{\refpa}[1]{(\ref{#1})} 
\nc{\calH}{\ensuremath{\mathcal{H}}}
\nc{\calD}{\ensuremath{\mathcal{D}}}
\nc{\calL}{\ensuremath{\mathcal{L}}}
\nc{\calO}{\ensuremath{\mathcal{O}}}
\nc{\hcalO}{\ensuremath{\hat \mathcal{O}}}
\nc{\calK}{\ensuremath{\mathcal{K}}}
\nc{\Tr}{\ensuremath{\mathrm{Tr}}}
\nc{\tr}{\ensuremath{\mathrm{tr}}}
\nc{\ra}{\rightarrow}
\nc{\lr}{\leftrightarrow}
\nc{\phistar}{\phi^*}
\nc{\etat}{\eta_T}
\nc{\het}{\hat E_T}
\nc{\hpt}{\hat \psi_T}
\nc{\hpdt}{\hat \psi ^\dagger_T}
\nc{\bart}{\bar{t}}
\nc{\barp}{\bar{p}}
\nc{\barT}{\bar{T}}
\nc{\hbarrho}{\hat{\bar{\rho}}}
\nc{\bga}{\ensuremath{\mbox{\boldmath{$\gamma$}}}}
\nc{\bsi}{\ensuremath{\mathbf{\sigma}}}
\nc{\bx}{\ensuremath{\mathbf{x}}}
\nc{\by}{\ensuremath{\mathbf{y}}}
\nc{\bz}{\ensuremath{\mathbf{z}}}
\nc{\bp}{\ensuremath{\mathbf{p}}}
\nc{\bn}{\ensuremath{\mathbf{n}}}
\nc{\bbp}{\ensuremath{\bar{\mathbf{p}}}}
\nc{\bP}{\ensuremath{\mathbf{P}}}
\nc{\hbA}{\hat{\ensuremath{\mathbf{A}}}}
\nc{\hbB}{\hat{\ensuremath{\mathbf{B}}}}
\nc{\bA}{\ensuremath{\mathbf{A}}}
\nc{\bJ}{\ensuremath{\mathbf{J}}}
\nc{\bB}{\ensuremath{\mathbf{B}}}
\nc{\bH}{\ensuremath{\mathbf{H}}}
\nc{\bM}{\ensuremath{\mathbf{M}}}
\nc{\bD}{\ensuremath{\mathbf{D}}}
\nc{\bE}{\ensuremath{\mathbf{E}}}
\nc{\hbE}{\hat{\ensuremath{\mathbf{E}}}}
\nc{\br}{\ensuremath{\mathbf{r}}}
\nc{\bj}{\ensuremath{\mathbf{j}}}
\nc{\bOm}{\ensuremath{\mathbf{\Om}}}
\nc{\om}{\omega}
\nc{\Om}{\Omega}
\nc{\sgn}{\mbox{sgn}}
\nc{\deltabar}{\mbox{$\delta\hspace*{-8pt}\vspace*{-8pt}-$}}
\nc{\gammat}{\tilde{\gamma}}
\nc{\mub}{\bar{\mu}}
\nc{\rhob}{\bar{\rho}}
\nc{\Bb}{\bar{B}}
\nc{\Jb}{\bar{J}}
\nc{\Mb}{\bar{M}}
\nc{\Tb}{\bar{T}}
\nc{\sbar}{\bar{s}}
\nc{\betab}{\bar{\beta}}
\nc{\hj}{\hat j}
\nc{\hQ}{\hat Q}
\nc{\hJ}{\hat J}
\nc{\hA}{\hat A}
\nc{\hH}{\hat H}
\nc{\de}{\delta}
\nc{\leri}{\leftrightarrow}
\nc{\llabel}[1]{\label{#1}\marginpar{#1}}
\nc{\bc}{\begin{center}}
\nc{\ec}{\end{center}}
\nc{\inv}[1]{\frac{1}{#1}}

\newlength{\overeqskip}
\newlength{\undereqskip}
\nc{\eq}[1]{\mbox{Eq.~(\ref{#1})}}
\nc{\eps}{\epsilon}
\nc{\goto}{\rightarrow}
\nc{\cF}{{\cal F}}
\nc{\cG}{{\cal G}}
\nc{\cH}{{\cal H}}
\newcounter{sectionc}
\newcounter{subsectionc}
\newcounter{subsubsectionc}
\renewcommand{\section}[1]
{\refstepcounter{sectionc}\vspace{0.0cm}
\setcounter{subsectionc}{0}\setcounter{subsubsectionc}{0}\noindent 
	{\bf\thesectionc. #1}}
\renewcommand{\subsection}[1] {\vspace{0.0cm}
\addtocounter{subsectionc}{1} 
	\setcounter{subsubsectionc}{0}\noindent 
	{\it\thesectionc.\thesubsectionc. #1}\par\vspace{0.4cm}}
\renewcommand{\subsubsection}[1] {\vspace{0.6cm}\addtocounter{subsubsectionc}{1}
	\noindent {\rm\thesectionc.\thesubsectionc.\thesubsubsectionc. 
	#1}\par\vspace{0.0cm}}

\renewcommand{\theequation}{\thesectionc.\arabic{equation}}
\renewenvironment{thebibliography}[1]
	{\begin{list}{\arabic{enumi}.}
	{\usecounter{enumi}\setlength{\parsep}{0pt}
\setlength{\leftmargin 1.25cm}{\rightmargin 0pt}
	 \setlength{\itemsep}{6pt} \settowidth
	{\labelwidth}{#1.}\sloppy}}{\end{list}}
\newcommand{\seqnoll}{\setcounter{equation}{0}}
\newcommand{\avg}[1]{\left\langle#1\right\rangle}
\begin{document}
%
%
%
%
%
\setlength{\jot}{10pt} 
%
%
\thispagestyle{empty} 
\vspace{1cm}
%
%
\begin{center}  
\baselineskip 1.2cm 
{\Huge\bf Causality  in  Quantum Field Theory 
with 
Classical Sources 
}
\\[5mm]  
\normalsize 
\end{center} 
{\centering 
{\large Bo-Sture K. Skagerstam\footnote{Corresponding author. Email address: bo-sture.skagerstam@ntnu.no}$^{,a)}$}, 
{\large Karl-Erik Eriksson\footnote{Email address: frtkee@chalmers.se}$^{,b)}$}, 
%
%
%
{\large Per K. Rekdal\footnote{Email address: per.k.rekdal@himolde.no}$^{,c)}$} 
\\[5mm] 
$^{a)}$Department of Physics, NTNU,  Norwegian University of Science and Technology, N-7491 Trondheim, Norway
\\[1mm]
$^{b)}$ Department of Space, Earth and Environment,
Chalmers University of Technology, SE-412 96 G\"{o}teborg, Sweden
\\[1mm]
$^{c)}$Molde University College, P.O. Box 2110,  N-6402 Molde, Norway
\\[1mm]
} 
%
%
%
%
%
%
%
%
%
%
%
\begin{abstract} 
%
%
\indent  
In an exact quantum-mechanical framework we show  
that space-time expectation values of the second-quantized electromagnetic fields  in the Coulomb gauge, in the presence of a classical source, automatically lead to {\it causal} and properly  {\sl retarded }  electromagnetic field strengths.   The classical $\hbar$-independent and gauge invariant Maxwell's equations then naturally emerge  and are therefore also  consistent with the classical special theory of relativity.  The fundamental difference between interference phenomena due to the linear nature of the classical Maxwell theory as considered in, e.g.,  classical optics,  and interference effects of quantum states is clarified.     In addition to these issues, the framework outlined also provides for  a simple approach to invariance under time-reversal, some  spontaneous  photon  emission and/or absorption processes as well as an approach  to   Vavilov-\v{C}herenkov radiation.  The inherent and necessary quantum uncertainty, limiting a precise space-time knowledge of expectation values of the quantum fields considered,  is, finally,  recalled. 
\vspace{1mm}
\end{abstract} 
%
%
%
%
%
%
\vspace{0.5cm}
\newpage
\setcounter{page}{1}
\seqnoll
\bc{
\section{\large Introduction}
\label{sec:introuction}
}\ec

The roles of causality and retardation in  classical and quantum-mechanical versions of electrodynamics are issues that one encounters in various contexts  (for recent  discussions see, e.g.,  Refs.\cite{brill_67}-\cite{skagerstam_2019}).  
%
%
In electrodynamics it is natural to introduce gauge-dependent scalar and vector potentials. These potentials do not have to be local  in space-time. 
It can then be a rather delicate issue to verify that  gauge-independent observables obey the physical constraint of causality and that they also are properly retarded. Attention to this and related issues are often discussed in a classical framework where one explicitly shows how  various choices of gauge  give rise to the same electromagnetic field strengths (see, e.g., the excellent discussion in Ref.\cite{jackson_02}). Even though issues related to  causality in physics have  been discussed for  many years, we are still facing new insights regarding such fundamental concepts.  In a recent investigation \cite{Budko_2009}  the near-, intermediate-, and far-field causal properties of classical electromagnetic fields have been discussed in great detail. In terms of experimental  and theoretical considerations,  locally backward velocities and apparent super-luminal features of electromagnetic fields   were demonstrated. Such observations do not challenge our understanding of causality since they describe phenomena that occur behind the light front of electromagnetic signals (see, e.g., Refs.\cite{lautrup_2001,Budko_2009,skagerstam_2019} and references cited therein).

In the present paper, it is our goal to recall the problems mentioned above in a quantum-mechanical framework. Some of theses aspects were  already considered a long time ago by Fermi \cite{Fermi_1932}.  We consider, in particular,  the {\it finite} and exact time-evolution as dictated by quantum mechanics with second-quantized electromagnetic fields in the presence of a general  classical conserved current.
In terms of suitable and well-known optical quadratures (see, e.g., Ref.\cite{schleich_2005}), the  corresponding $\hbar$-dependent dynamical equations can then be reduced to a system of decoupled harmonic oscillators with a space-time dependent external force. 
No pre-defined global causal order is assumed other than the deterministic time-evolution as prescribed by the Schr\"{o}dinger equation. The classical $\hbar$-independent theory of Maxwell then naturally emerges  in terms of  properly causal and retarded  expectation values of  the second-quantized electromagnetic field for any initial quantum state. This is in line with more general $S$-matrix arguments due to Weinberg \cite{weinberg_1965}.  
Furthermore,  the fundamental role of interference in the sense of quantum mechanics as compared to classical interference effects due to the linearity of Maxwell's equations, can then be clarified. The quantum-mechanical approach also leads to a deeper insight with regard to the role of unavoidable quantum uncertainty of average values of quantum fields. 

This presentation can, in a rather straightforward manner, be extended to   gravitational quantum uncertainties around a flat Minkowski space-time,  in the presence of a classical source in terms of a conserved energy-momentum tensor. As a result,  the classical weak-field limit of Einstein's theory of general relativity emerges. This is discussed in a separate publication \cite{ESR_II}.

   The paper  is organized as follows. In Section  \ref{sec:Maxwell} we recall, for reasons of completeness,  the classical version of electrodynamics in vacuum and the corresponding issues of causality and retardation  in the presence of a space-time dependent source, and the extraction of   a proper set of physical but non-local degrees of freedom. The exact quantum-mechanical framework approach is  illustrated in terms of a   second-quantized single-mode electromagnetic field in the presence of a time-dependent classical source  in Section \ref{sec:one_mode_only}, where emergence of the classical $\hbar$-independent physics is also made explicit.  In Section \ref{sec:multimode}, the analysis of Section \ref{sec:one_mode_only} is extended to multi-modes and to a general space-time dependent classical source. The issues of causality, retardation, and time-reversal are then discussed in Section \ref{sec:causality}. The framework also provides for a discussion of some radiative processes, and in  Section \ref{sec:gphoton_emission_proc} we  consider dipole radiation, and  the fameous classical Vavilov-\v{C}herenkov radiation is reproduced in a straightforward and exact manner. In Section \ref{sec:fluctuations}, we briefly discuss the role of the  intrinsic quantum uncertainty of expectation values considered.   Finally,  and in Section \ref{sec:final_remarks}, we present conclusions and final remarks.  Some multi-mode considerations as referred to in the main text are presented in an Appendix.
\newline
%
%

\vspace{0.5cm}
\bc{
\section{\large Maxwell's Equations with  a Classical Source}
\label{sec:Maxwell}
}\ec
%
%
%

 Unless stated explicitly, we often make use of the  notation ${\bf E} \equiv {\bf E}({\bf x},t)$ for the electric field and similarly for other fields. The microscopic classical Maxwell's equations in vacuum are then (see, e.g., Ref.\cite{Jackson75}):
\begin{gather}  \label{eq:causality1}
   \nabla \cdot {\bf E} = \frac{\rho}{\epsilon_0} ~ , 
   \\ \label{eq:causality2}
   \nabla \cdot {\bf B} = 0 ~ , 
   \\  \label{eq:causality3}
   \nabla \times {\bf E} = - \, \frac{\partial {\bf B}}{\partial t} ~ , 
   \\  \label{eq:causality4}
   \nabla \times {\bf B} = \, \mu_0 \, {\bf j} + \epsilon_0\mu_0 \, \frac{\partial {\bf E}}{\partial t} ~ , 
\end{gather}
with the velocity of light in vacuum as given by $c=1/\sqrt{\epsilon_0\mu_0}$.
Eqs.(\ref{eq:causality1}) and (\ref{eq:causality4}) imply current conservation, i.e., 
\begin{equation}
\label{eq:conservation}
    \frac{\partial \rho}{\partial t} + \nabla \cdot {\bf j} = 0 ~ .
\end{equation}
The classical Maxwell's equations can, of course, be written in a form that is  explicitly covariant under Lorentz transformations but this will not be of importance here.
 
 The general vector identity
\begin{equation} 
\label{eq:identity}
     \nabla \times ( \nabla \times {\bf F}) = \nabla ( \nabla \cdot {\bf F}  )
- \nabla^2 {\bf F}  ~ ,
\end{equation}
%
%
applied to the electric field ${\bf E}$ and making use of  Maxwell's equations (\ref{eq:causality3}) and (\ref{eq:causality4}) implies that
\begin{equation} \label{eq:causality6}
   \frac{1}{c^2} \, \frac{\partial ^2 {\bf E}}{\partial t ^2} - \nabla^2 {\bf E} =  - \mu_0 \, \frac{\partial \, {\bf j}}{\partial t} - \nabla \bigg ( \frac{\rho}{\epsilon_0} \bigg ) ~ ,
\end{equation}
with retarded as well  as advanced solutions. By physical arguments one  selects the retarded solution, even though Maxwell's equations are invariant under time-reversal as, e.g.,  discussed by Rohrlich \cite{rohrlich_01}.

 We now write the electric field ${\bf E}$ and the magnetic field ${\bf B}$ in terms of the vector potential ${\bf A}$ and the scalar potential $\phi$, i.e.,
\begin{equation}
    {\bf E} = - \frac{\partial {\bf A}}{\partial t} - \nabla \phi ~ ,
\end{equation}
and
\begin{equation}
    {\bf B} = \nabla \times {\bf A} ~ .
\end{equation}
The  Coulomb (or radiation) gauge, which, of course,  is not Lorentz covariant,  is defined by the requirement
\begin{equation}
   \nabla \cdot {\bf A} = 0 ~ ,
\end{equation} 
and therefore leads to at most two physical degrees of freedom of the electromagnetic field. ${\bf A}$ defined with this gauge-choice restriction is denoted by ${\bf A}_T$. 
By making use of the vector identity Eq.(\ref{eq:identity}) with ${\bf F}={\bf A}_T$, 
Amp$\grave{\textrm{e}}$re's law, i.e., Eq.(\ref{eq:causality3}), may then be written in the form
\begin{equation} \label{eq:causality5}
   \frac{\partial^2 {\bf A}_T}{\partial t^2} - c^2 \nabla^2 {\bf A}_T = \frac{{\bf j}_T}{\epsilon_0} ~ ,
\end{equation}
where we have introduced a transverse current ${\bf j}_T$ according to
\begin{equation}
     {\bf j}_T \equiv {\bf j} - \epsilon_0 \frac{\partial}{\partial t} \nabla \phi ~ .
\end{equation}
Eq.(\ref{eq:causality5}) is, of course, the well-known wave-equation for the vector potential ${\bf A}_T$ in the Coulomb gauge.
The transversality condition  $ \nabla \cdot {\bf j}_T = 0$  follows from charge conservation and 
\begin{equation}
\label{eq:laplace}
    \nabla \cdot {\bf E} = - \nabla^2 \phi = \frac{\rho}{\epsilon_0} ~ ,
\end{equation}
 in the Coulomb gauge. Eq.(\ref{eq:laplace}) is, therefore, not dynamical but should rather be regarded as a constraint on the physical degrees of freedom in the Coulomb gauge enforcing current conservation. The instantaneous scalar potential $\phi$ degree of freedom can therefore be  eliminated entirely in terms of the physical charge density $\rho$ (in this context see, e.g.,  Refs.\cite{Sakurai_1967,cohen_1989}).

In passing we also recall that in the Coulomb gauge, the scalar potential $\phi$ is, according to Eq.(\ref{eq:laplace}), given by
\begin{equation}
    \phi({\bf x},t) = \frac{1}{4\pi \epsilon_0} \, \int d^3x' \, \frac{\rho({\bf x}',t)}{|{\bf x}-{\bf x}'|} ~ .
\end{equation}
Due to the conservation of the current, i.e., Eq.(\ref{eq:conservation}), the time derivative of $\phi$ may be written in the form
\begin{equation}
    \frac{\partial \phi({\bf x},t)}{\partial t} = - \frac{1}{4\pi \epsilon_0} \, \int d^3x' \, \frac{\nabla '\cdot {\bf j}({\bf x}',t)}{|{\bf x}-{\bf x}'|} ~ .
\end{equation}
According to the well-known Helmholtz decomposition  theorem ${\bf F} = {\bf F}_L + {\bf F}_T$ for a vector field (see, e.g.,  Ref.\cite{Jackson75}), formally written in the form
\begin{equation} \label{eq:causality11}
  {\bf F} = \frac{1}{\nabla^2} \, \bigg ( \, \nabla ( \nabla \cdot {\bf F} ) - \nabla\times(\nabla\times{\bf F}) \bigg ) ~ , 
\end{equation}
%
%
using Eq.(\ref{eq:identity}),  we can identify the corresponding longitudinal current ${\bf j}_L$, i.e.,
\begin{equation}  \label{eq:causality13}
    {\bf j}_L({\bf x},t) \equiv -\frac{1}{4\pi}{\nabla } \int d^{3}x' \frac{\nabla ' \cdot {\bf j}({\bf x}',t)}{| {\bf x} - {\bf x}' |} ~ .
\end{equation}
%
%
It is now evident that the right-hand side of the wave-equation Eq.(\ref{eq:causality5}) for the vector potential  can be expressed in terms of  the current ${\bf j}({\bf x},t)$, i.e., 
\begin{equation} 
    {\bf j}_T({\bf x},t) \equiv \frac{1}{4\pi} {\nabla}{\times} ({\nabla}{\times}{\int d^{3}x' \frac{{\bf j}({\bf x}',t)}{|{\bf x} - {\bf x}'|}} )\,\,  .
\end{equation}
The important point here is  that ${\bf j}_T$  is  an instantaneous and non-local function in space of the physical current ${\bf j}({\bf x},t)$. When the Helmholtz decomposition  theorem is applied to the vector potential ${\bf A}= {\bf A}_L + {\bf A}_T$, it follows that the transverse part ${\bf A}_T$ is gauge-invariant but, again,  a non-local function in space of  the vector potential ${\bf A}$.
%
%
%

 At the classical level, we now make a normal-mode Ansatz for the real-valued vector field ${\bf A}$ confined in, e.g., a  cubic box with volume $V=L^3$ and with periodic boundary conditions. 
With ${\bf k} =2\pi (n_x,n_y,n_z)/L$, where  $n_x,n_y,n_z$ are integers, we  therefore write
\begin{equation} 
\label{classical_field}
  {\bf A}_T({\bf x},t) = \sum_{{\bf k}\lambda} \,  \sqrt{ \frac{1}{V \epsilon_0} } \,  \,
 \bigg ( \,  q_{{\bf k} \lambda}(t) {\bm{\epsilon}}({\bf k} ;\lambda)e^{i {\bf k} \cdot {\bf x}} + q^*_{{\bf k} \lambda}(t){\bm{\epsilon}}^*({\bf k} ;\lambda)  e^{-i {\bf k} \cdot 
{\bf x}} \, \bigg ) ~ ,
\end{equation}
with time-dependent Fourier components $q_{{\bf k} \lambda}(t)$.  The, in general,  complex-valued polarization vectors ${\bm{\epsilon}}({\bf k} ;\lambda)$ obey the transversality condition ${\bf k} \cdot  {\bm{\epsilon}}({\bf k} ;\lambda) = 0$.
 They are normalized in such a way that 
\begin{equation}
\label{eq:polarization_sum}
P_{ij} \equiv  P_{ij}(\hat{\bf k}) \equiv \sum_{\lambda }{{\epsilon}}^*_i({\bf k} ;\lambda){{\epsilon}}_j({\bf k} ;\lambda) = \delta_{ij} - \hat{{ k}}_i\hat{{ k}}_j \, \, ,
\end{equation}
where we have defined the unit vector $\hat{{\bf k}} \equiv {\bf k}/|{\bf k}|$. 
In the case of linear polarization the real-valued, orthonormal, and linear polarization  unit vectors ${\bm{\epsilon}}({\bf k} ;\lambda)$, with $\lambda = 1,2$,  are such that  ${\epsilon}_i(-{\bf k} ;\lambda) = (-1)^{\lambda +1}{\epsilon}_i({\bf k} ;\lambda)$. Since ${\bf A}_T$ itself is independent of the actual realization of the polarization degrees of freedom ${\bm{\epsilon}}({\bf k} ;\lambda)$,  it is without any difficulty to express Eq.(\ref{classical_field}) in terms of, e.g.,  the complex circular polarization vectors  with $\lambda = \pm$, i.e., 
\begin{equation}
\label{eq:circular_polarization}
{\bm{\epsilon}}({\bf k};\pm)= \frac{1}{\sqrt{2}}\big({\bm{\epsilon}}({\bf k} ;1) \pm i{\bm{\epsilon}}({\bf k} ;2)\big)\,\, , 
\end{equation}
such that ${\bm{\epsilon}}(-{\bf k};\pm)= {\bm{\epsilon}}^*({\bf k};\pm)$. 

The Ansatz Eq.(\ref{classical_field}) for ${\bf A}_T$ is, of course, consistent with transversality of the current ${\bf j}_T$ in Eq.(\ref{eq:causality5}). Due to the transversality of ${\bf j}_T$,  we can then also write that
\begin{equation} 
  {\bf j}_T({\bf x},t) = 
\sum_{{\bf k}\lambda} \,  \sqrt{ \frac{\epsilon_0}{V } } \,  \bigg ( \, j_{{\bf k} \lambda}(t) {\bm{\epsilon}}({\bf k} ;\lambda) e^{i {\bf k} \cdot {\bf x}} + j^*_{{\bf k} \lambda}(t) {\bm{\epsilon}}^*({\bf k} ;\lambda) e^{-i {\bf k} \cdot {\bf x}} \, \bigg ) ~ .
\end{equation}
\newline
\noindent The time-dependence of  $q_{{\bf k}\lambda}(t)$ is now determined by the dynamical equation Eq.(\ref{eq:causality5}) for ${\bf A}_T$, i.e.,
\begin{equation}
     \ddot{q}_{{\bf k} \lambda}(t) + \omega_k^2 \, q_{{\bf k} \lambda}(t) =  j_{{\bf k}\lambda}(t) ~  ,
\end{equation}
with $\omega_k = c|{\bf k}|$. If we define classical real-valued quadratures
\begin{equation}
    Q_{{\bf k} \lambda}(t) \equiv  q_{{\bf k} \lambda}(t) + q^*_{{\bf k} \lambda}(t) ~ ,
\end{equation}
then
\begin{equation}
\label{eq:quadrature}
    \ddot{Q}_{{\bf k} \lambda}(t) + \omega_k^2 \, Q_{{\bf k} \lambda}(t) =  j_{{\bf k} \lambda}(t) + j^*_{{\bf k} \lambda}(t) \equiv f_{{\bf k} \lambda}(t) ~ .
\end{equation}
This equation has the same  form as the dynamical equation for a time-dependent forced harmonic oscillator. The corresponding quantum dynamics will be treated in the next session.
\newpage
%
\vspace{0.2cm}
\bc{
\section{\large Single Mode Considerations}
\label{sec:one_mode_only}
}
\ec
\seqnoll

As seen in the previous section, a single-mode of the electromagnetic field reduces to a dynamical system equivalent to  a forced harmonic oscillator with a time-dependent external force. The quantization of such a system is well-known (see, e.g., Refs.\cite{Feynman_1951}-\cite{Knight_1977}) and is presented here in a  form suitable for  illustrating a calculational procedure to be used in later sections for {\it finite} time intervals.  

With only one mode present,  we write $Q \equiv Q_{{\bf k} \lambda}(t)$, $\omega \equiv \omega_k$, as well as $f(t) \equiv  f_{{\bf k} \lambda}(t)$.    Eq.(\ref{eq:quadrature}) then takes the form
\begin{equation}
\label{eq:3.1}
    \ddot{Q} + \omega^2 Q = f(t) ~ .
\end{equation}
\noindent This classical equation of motion can, of course,  be obtained from the classical time-dependent Hamiltonian $H_{cl}(t)$  for a forced harmonic oscillator  with $m=1$, i.e.,
\begin{equation}
  H_{cl}(t) = \frac{P^2}{2} + \frac{1}{2} \omega^2 Q^2 - f(t)Q ~ .
\end{equation}
\noindent We  quantize this classical system by making use of  the canonical commutation relation
\begin{equation}
    \big [ \, Q,P \, \big ] = i \hbar ~ . 
\end{equation}
We express $Q$ and $P$ in terms of the quantum-mechanical quadratures
\begin{equation}
\label{eq:Q_quadrature}
     Q = \sqrt{ \frac{\hbar}{2 \omega} } \, \big ( a + a^* \big ) ~ , 
\end{equation}
as well as
\begin{equation}
\label{eq:P_quadrature}
     P = i \sqrt{ \frac{\hbar \omega}{2} } \, \big ( a^* - a \big ) ~ , 
\end{equation}
where  $[a,a^*]= 1$. The classical Hamiltonian $H_{cl}(t)$ is then promoted to the explicitly  time-dependent quantum-mechanical  Hamiltonian $H(t)$ according to
\begin{equation}
\label{eq:hamiltonian}
   H_{cl}(t) \rightarrow H(t) = \hbar \omega \, \Big ( a^*a + \frac{1}{2} \Big ) + g(t) \Big ( a + a^* \Big ) ~ ,
\end{equation}
where we have defined  
\begin{equation}
    g(t) \equiv - f(t) \sqrt{ \frac{\hbar}{2\omega} } ~ .
\end{equation}

In general,  it is notoriously difficult to solve   the Schr\"{o}dinger equation with an explicitly time-dependent Hamiltonian. Due to the at most quadratic dependence of $a$ and $a^*$  in Eq.(\ref{eq:hamiltonian}) it is, however, easy to solve exactly  for the unitary quantum dynamics.
Indeed, if one considers the dynamical evolution of the system in the interaction picture with $\ket{\psi(t)}_I \equiv \exp(itH_0/\hbar)\ket{\psi(t)}$, where we for convenience make the choice $t_0 =0$ of initial time, then 
\begin{equation}
     i \hbar \, \frac{d\ket{\psi(t)}_I}{dt} = H_I(t) \ket{\psi(t)}_I ~ .
\end{equation}
For observables  ${\cal O}$ in the interaction picture we also have that
\begin{equation}
     {\cal O}_I(t) \equiv \exp(it H_0/\hbar) {\cal O}\exp(-it H_0/\hbar) ~ .
\end{equation}
In our case $H_0= \hbar\omega (a^*a+1/2)$ and therefore

\begin{equation}
\label{eq:onemode_interaction}
      H_I(t) = g(t)\left( ae^{-i\omega t} + a^{*}e^{i\omega t} \right) ~ .
\end{equation}
 The explicit solution for $\ket{\psi(t)}_I$ is then given by
\begin{equation}
\label{eq:timeevolution}
     \ket{\psi(t)}_I = \exp \left ( \frac{i}{\hbar} \phi(t) \right ) \, \exp \left ( - \frac{i}{\hbar} \int_{0}^t dt' \, H_I(t') \right ) \, \ket{\psi(0)} ~ ,     
\end{equation}
for any initial pure state $\ket{\psi(0)}$. Eq.(\ref{eq:timeevolution}) can easily be verified by, e.g.,  considering the limit $\Delta t \rightarrow 0$ of $(\ket{\psi(t+\Delta t)}_I-  \ket{\psi(t)}_I)/\Delta t$ using that $\exp(A+B)=\exp(A)\exp(B)\exp(-[A,B]/2)$ if $[A,B]$ is a c-number. The $c$-number phase $\phi(t)$ can then be computed according to
\begin{equation}
    i \phi(t) = \frac{1}{2\hbar}  \int_{0}^t  dt'  \Big [ N(t') , H_I(t') \Big ]  = \frac{1}{2\hbar}  \int_{0}^t dt'\int_{0}^{t'} dt''  \Big [ H_I(t'') , H_I(t') \Big ]\,\,,
\end{equation}
with 
\begin{equation}
      N(t) \equiv \int_0^t dt' \, H_I(t') ~ ,
\end{equation}
since, in our case, $[ \, N(t') , H_I(t') \, ]$ is a $c$-number.
 We therefore see that, apart from a phase, the time-evolution in the interaction picture is controlled by a conventional displacement operator as used in  various studies of coherent states (see, e.g., Refs.\cite{Klauder_Skagerstam,Skagerstam_Klauder} and references cited therein).

The expectation value of the quantum-mechanical quadrature $Q$ in Eq.(\ref{eq:Q_quadrature}) at  time $t$, i.e., ${\avg{Q}(t)}\equiv \bra{\psi(t)}Q \ket{\psi(t)} ={_I}\bra{\psi(t)}{Q_I(t)}\ket{\psi(t)}_I$,
can now easily be evaluated for an arbitrary  initial pure state $\ket{\psi(0)}$ with the result
\begin{eqnarray}
\label{classical_int_23}
{\avg{Q}}(t) = {\avg{Q}} \cos(\omega t) +\frac{1}{\omega}{\avg{P}} \sin (\omega t) +\frac{1}{\omega}\int_{0}^{t}dt'f(t')\sin \Big(\omega(t - t')\Big) \, ,~~~~~~~~~
\end{eqnarray}
where ${\avg{{\cal O}}} \equiv {\avg{{\cal O}}}(0)$ for the initial expectation value of an observable ${\cal O}$.
In Eq.(\ref{classical_int_23})  we, of course,  recognize the general classical solution of the forced harmonic oscillator equations of motion
Eq.(\ref{eq:3.1}), i.e., 
\begin{eqnarray}
\label{classical_equations}
\frac{d^2}{dt^2}{\avg{Q}}(t) +\omega^2 {\avg{Q}}(t) = f(t)\, ,~~~~~~~~~
\end{eqnarray}
 in terms of its properly  retarded Green's function (see, e.g., Ref.\cite{Byron_Fuller_1992}). The last term in Eq.(\ref{classical_int_23}) is classical in the sense that it does not depend on $\hbar$. Possible quantum-interference effects are hidden in the homogeneous solution of Eq.(\ref{classical_equations}).
 Similarly, we find for the $P$-quadrature in Eq.(\ref{eq:P_quadrature}) that ${\avg{P}}(t) = d{\avg{Q}}(t)/dt$ or more explicitly:
\begin{eqnarray}
\label{classical_int_24}
{\avg{P}}(t) = {\avg{P}} \cos (\omega t) -{\avg{Q}} \omega \sin(\omega t)  +\int_{0}^{t}dt'f(t')\cos \Big(\omega(t - t')\Big) \, .~~~~~~~~~
\end{eqnarray}

Even though the classical equation of motion emerges in terms of quantum-mechanical expectation values, intrinsic quantum uncertainty for any observable ${\cal O}$ as defined by
\begin{eqnarray}
\label{classical_int_24_2}
(\Delta {\cal O})^2(t) \equiv \bra{\psi(t)}({\cal O} - {\avg{\cal O}})^2 \ket{\psi(t)} \,  ,~~~~~~~~~
\end{eqnarray}
are in general  present. For ${\cal O}=Q$ one finds that
\begin{eqnarray}
\label{classical_int_24_3}
(\Delta Q)^2(t) = (\Delta Q )^2\cos^2\omega t+ (\Delta P)^2\frac{\sin^2\omega t}{\omega ^2}
+ \frac{\cos \omega t \sin \omega t}{\omega}{\avg{PQ+QP -2{\avg{Q}}{\avg{P}}}}\,\, ,~~~~
\end{eqnarray}
independent of the external force $f(t)$. For minimal dispersion states, i.e., states for which $\Delta Q\Delta P=\hbar /2$, the last term is zero. For coherent states one then finds the intrinsic and time-independent quantum-mechanical uncertainty  $(\Delta Q )^2(t) = \hbar/2\omega$ and 
$(\Delta P)^2(t) = \hbar\omega/2$.

The classical equation of motion Eq.(\ref{classical_equations}) allows for linear superpositions of solutions. Such linear superposition are, however, not directly related to quantum-mechanical superpositions of the initial quantum states since expectation values are non-linear functions of quantum states.  For number states $\ket{n}\equiv (a^{*})^n\ket{0}/\sqrt{n!}$, with   $a\ket{0} = 0$, which  in terms of, e.g., a Wigner function have no classical interpretation except for the vacuum state $\ket{0}$ (see, e.g., Ref.\cite{schleich_2005}), we have that 
${\avg{Q}}= {\avg{P}} = 0$ but $(\Delta Q )^2(t) =(\Delta P )^2(t)/\omega^2 = (n+1/2)\hbar/\omega$. For an initial state of the form $\ket{\psi (0)} =(\ket{0}+\ket{1})/\sqrt{2}$ we find that ${\avg{Q}}= \sqrt{\hbar/2\omega}$ and ${\avg{P}} = 0$ with 
 an intrinsic time-dependent quantum uncertainty, e.g., $(\Delta Q )^2(t) =\hbar(2-\cos^2\omega t)/2\omega$. This initial state therefore leads to  expectation values that do not correspond to a superposition of the classical solutions  obtained from the initial states $\ket{0}$ or $\ket{1}$. This simple example demonstrates the fundamental difference between the role of the superposition principle in classical and in quantum physics. It is a remarkable achievement of experimental quantum optics that such  quantum-mechanical interference effects between the vacuum state and  a single-photon  state have been observed \cite{Ou_90,Mandel_99} (for a related discussions also see Ref.\cite{Resch_2002}-\cite{Johansson_2016}). In the next section we extend this simple single-mode case to the general multi-mode space-time dependent  situation.
%
%
%
%
%
\vspace{0.2cm}
\bc{
\section{Multi-Mode Considerations}
\label{sec:multimode}
}\ec
\seqnoll

\begin{figure}[htb]  
\vspace{-2cm}
\centerline{\includegraphics[width=12cm,angle=0]{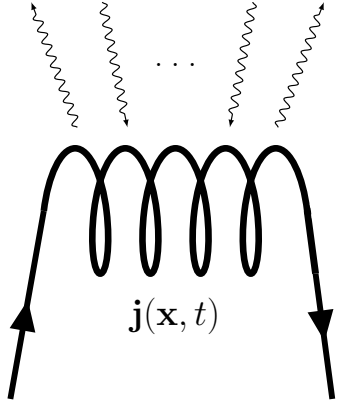}}
\vspace{-7cm}                 
\caption{Absorption and emission of photons from a classical current ${\bf j}({\bf x},t)$.}
\label{fig:classical_1}
%
\end{figure}

We will now consider emission as well as absorption processes of photons in the presence of a general space-time dependent classical source as illustrated in Fig.\ref{fig:classical_1}.
%
In the multi-mode case  the interaction Hamiltonian $H_I(t)$ for a classical current ${\bf j}$ is now an extension of the single-mode version Eq.(\ref{eq:onemode_interaction}). In the Coulomb gauge and  in the interaction picture,  we therefore consider
\begin{gather} 
  H_I(t)  = - \int_V d^3x ~ {\bf j}({\bf x},t) \cdot {\bf A}_T({\bf x},t)
 \nonumber \\
  = - \sum_{{\bf k} \lambda} \, \sqrt{ \frac{\hbar}{2V\epsilon_0 \omega_k} } \,  \, \bigg ( \,{\bf j}({\bf k},t)\cdot{\bm \epsilon}({\bf k} ;\lambda)  a_{{\bf k} \lambda}e^{- i \omega_k t}  + {\bf j}^*({\bf k},t)\cdot{\bm \epsilon}^*({\bf k} ;\lambda) a^*_{{\bf k} \lambda} e^{i \omega_k t}  \, \bigg ) ~ ,
\label{eq:multimode_1}
\end{gather}
with the ``free'' field Hamiltonian
\begin{gather} 
H_0 = \sum_{{\bf k}\lambda}\hbar\omega_k(a^*_{{\bf k} \lambda} a_{{\bf k} \lambda}+ \frac{1}{2}) \,\, .
\label{eq:multimode_2}
\end{gather}
 Here we have introduced the Fourier transformed current 
\begin{equation}
     {\bf j}({\bf k},t) \equiv \int_V d^3x ~ e^{i {\bf k} \cdot {\bf x}} \, {\bf j}({\bf x},t) ~ .
\end{equation}
Since ${\bf j}_L({\bf k},t)= \hat{{\bf k}} (\hat{{\bf k}}\cdot{\bf j}({\bf k},t))$ and ${\bf j}_T({\bf k},t)= {\bf j}({\bf k},t)- {\bf j}_L({\bf k},t)$, it is clear due to the transversality condition ${\bf k} \cdot  {\bm{\epsilon}}({\bf k} ;\lambda) = 0$ that  only the transverse part of the current  contributes  in Eq.(\ref{eq:multimode_1}).

The interaction Eq.(\ref{eq:multimode_1}) therefore corresponds to a system of independent forced harmonic oscillators  of the one-mode form as discussed in the previous section.
The second-quantized version of the vector-potential  in  Eq.(\ref{classical_field}) then has  the form of a free quantum field, i.e.,
\begin{equation} 
\label{eq:em_quantized}
  {\bf A}_T({\bf x},t) = \sum_{{\bf k}\lambda} \,  \sqrt{ \frac{\hbar}{2V \epsilon_0 \omega_k} } \, \,
 \bigg ( \, a_{{\bf k} \lambda}{\bm \epsilon}({\bf k};\lambda)  e^{i( {\bf k} \cdot {\bf x}- \omega_k t)} + a^*_{{\bf k} \lambda} {\bm \epsilon}^*({\bf k};\lambda)  e^{-i ({\bf k} \cdot 
{\bf x}-\omega_k t)} \, \bigg ) ~ ,
\end{equation}
with the basic canonical commutation relation 
\begin{equation} 
\label{eq:em_ccr}
[a_{{\bf k} \lambda} ,a^*_{{\bf k'} \lambda'} ] = \delta_{\lambda \lambda'} \delta_{{\bf k}{\bf k'}}\,\, ,
\end{equation}
 and where we recall that $\omega_k = c|{\bf k}|$. The vacuum state $|0 \rangle$ is then such that $a_{{\bf k} \lambda} |0 \rangle = 0$ for all quantum numbers ${\bf k} \lambda$.  The quantum field ${\bf A}_T$ is then  normalized in such a way that
\begin{equation} 
\label{eq:free_hamiltonian}
H_0 =\int_V d^3x\frac{1}{2}\left(\varepsilon_0{\bf E}_T^2({\bf x},t) +\frac{1}{\mu_0}{\bf B}^2({\bf x},t)\right)\,\, ,
\end{equation}
where we, for the free field in Eq.(\ref{eq:em_quantized}), make use of ${\bf E}_T= -\partial {\bf A}_T/\partial t$ and ${\bf B} =\nabla \times{\bf A} =\nabla \times{\bf A}_T$. 

If we consider the circular polarization vectors ${\bm \epsilon}({\bf k};\pm)$ according to Eq.(\ref{eq:circular_polarization}), we have to replace the annihilation operators $a_{{\bf k} \lambda}$ with
\begin{equation}
\label{eq:helicity_basis}
a_{{\bf k} \pm }= \frac{1}{\sqrt{2}}(a_{{\bf k} 1} \mp i a_{{\bf k}2})\,\, . 
\end{equation}
We  then observe  that
\begin{equation}
\label{eq:change_of_basis}
\sum_{\lambda =1,2} {\bm \epsilon}({\bf k};\lambda)a_{{\bf k}\lambda} = \sum_{\lambda =\pm} {\bm \epsilon}({\bf k};\lambda)a_{{\bf k}\lambda}\,\, ,
\end{equation}
which means that the quantum field ${\bf A}_T$ does not depend on the actual realization of the choice of polarization degrees of freedom.

The  single photon quantum states $|{\bf k}\lambda \rangle \equiv a^*_{{\bf k} \lambda }|0\rangle$, with $\lambda =\pm$, will then  carry the energy $\hbar \omega_k$,  momentum $\hbar {\bf k}$ as well as the intrinsic spin angular momentum  $\pm \hbar$ along the direction $\hat{{\bf k}}$, i.e., the helicity quantum number of a massless spin-one particle. In passing, we  remark that the latter property can be inferred from a  consideration of   a rotation with an angle $\theta$ around the wave-vector ${\bf k}$ in terms of a rotation matrix $R_{ij}(\theta)$, which implies that $a_{{\bf k} \pm } \rightarrow a_{{\bf k} \pm }(\theta) =\exp(\pm i\theta)a_{{\bf k} \pm }$. In terms of the corresponding rotated polarization vectors ${\epsilon}_i({\bf k};\lambda|\theta)= R_{ij}(\theta){\epsilon}_j({\bf k};\lambda)$ we then have, in accordance with Eq.(\ref{eq:change_of_basis}),   that
\begin{equation}
\label{eq:change_of_basis_2}
\sum_{\lambda =1,2} {\bm \epsilon}({\bf k};\lambda)a_{{\bf k}\lambda} = \sum_{\lambda =\pm} {\bm \epsilon}({\bf k};\lambda|\theta)a_{{\bf k}\lambda}(\theta)\,\, .
\end{equation}
In addition to the intrinsic spin angular momentum, photon states can also carry conventional orbital  angular momentum which plays an important role in many current contexts (see, e.g., Ref.\cite{OAM_2003} and references cited therein) but will not be of concern in the present work. A complete set of physical and well-defined Fock-states can then be generated in a conventional manner. By construction, these states have positive norm avoiding the presence of indefinite norm states in manifestly covariant formulations (for some considerations see, e.g., Refs.\cite{Lautrup_1967,Nakanishi_1972,Futakuchi_2017}).

Since, obviously,
\begin{gather} \nonumber
  \int_0^t dt' \, H_I(t') =~~~~~~
  \\
\label{eq:multimode_H}
   - \sum_{{\bf k} \lambda} \, \sqrt{ \frac{\hbar}{2V\epsilon_0 \omega_k} } \, \int_0^t dt'  \bigg (  a_{{\bf k} \lambda} \,  e^{- i \omega_k t'} {\bm \epsilon}({{\bf k} ;\lambda})\cdot {\bf j}({\bf k},t') + a^*_{{\bf k} \lambda} \,  e^{i \omega_k t'} {\bm \epsilon}^*({{\bf k} ;\lambda})\cdot {\bf j}^*({\bf k},t')  \bigg ) ~ ,
\end{gather}
we conclude that the time-evolution for $\ket{\psi(t)}_I$ in Eq.(\ref{eq:timeevolution}) is, apart from a phase factor, given by a multi-mode displacement operator
\begin{equation}
\label{eq:def_of_D}
  D({\bm \alpha}) \equiv  \exp \left ( - \frac{i}{\hbar} \int_{0}^t dt' \, H_I(t') \right )  = \prod_{{\bf k}\lambda} \, \exp \Big ( \alpha_{{\bf k}\lambda}(t) a^*_{{\bf k}\lambda} - {\alpha}^*_{{\bf k}\lambda}(t) a_{{\bf k}\lambda} \Big ) ~ .
\end{equation}
Here  $\alpha_{{\bf k}\lambda}(t)$ is, as  inferred from Eq.(\ref{eq:multimode_H}), explicitly  given by
\begin{gather} 
   \alpha_{{\bf k}\lambda}(t) \equiv \frac{i}{\hbar} \, \sqrt{ \frac{\hbar}{2V\epsilon_0 \omega_k} } \, \int_0^t dt' \, e^{i \omega_k t'} {\bf j}^*({\bf k},t') \cdot {\bm \epsilon}^*({\bf k}; \lambda) \,\, ,
\label{eq:alpha}
\end{gather}
with ${\bf j}^*({\bf k},t)={\bf j}(-{\bf k},t)$. The displacement operator $D({\bm \alpha})$ 
 has the form of a product of independent single-mode displacement operators.  By making use of Eq.(\ref{eq:timeevolution}), and by considering the action on the vacuum state, the quantum-mechanical time-evolution generates  a multi-mode coherent state $ D({\bm \alpha})\ket{0}$, apart from  the $\hbar$-dependent phase  $\phi (t)$ in Eq.(\ref{eq:timeevolution}).  As in the single-mode case, the time-dependent expectation value of the transverse quantum field ${\bf A}_T({\bf x},t)$ will then obey a classical equation of motion like Eq.(\ref{classical_equations}), i.e.,  (see Appendix A)
\begin{equation}
\label{eq:average_motion}
    \frac{\partial^2 \langle {\bf A}_T({\bf x},t) \rangle }{\partial t^2} - c^2 \nabla^2 \langle {\bf A}_T({\bf x},t) \rangle = \frac{{\bf j}_T({\bf x},t)}{\epsilon_0} ~ ,
\end{equation}
 to be investigated in more detail in Section \ref{sec:causality}. 
In other words, 
 there are particular quantum states of the radiation field, namely multi-mode coherent states,  which naturally lead to the classical electromagnetic fields obeying Maxwell's equations Eqs.(\ref{eq:causality1})-(\ref{eq:causality4}) in terms of quantum-mechancial expectation values. 
%
%
%
%
%
%
%
\vspace{0.5cm}
%
\bc{
\section{The Causality Issue}
\label{sec:causality}
}\ec
\seqnoll

 The expectation value of the transverse second-quantized vector field ${\bf A}_T $ is now given by Eq.(\ref{eq:causality19}), i.e.,
\begin{gather}  
    \langle {\bf A}_T({\bf x},t) \rangle =   \frac{1}{\epsilon_0} \int \frac{d^3k}{(2\pi)^3} e^{ i {\bf k} \cdot {\bf x} } \int_0^t dt' \frac{ \sin \big ( \omega_k (t-t') \big ) }{\omega_k} 
\int d^3x' e^{-i{\bf k} \cdot {\bf x}'} {\bf j}_T({\bf x}',t') ~ ,
\label{eq:meanfield}
\end{gather}
where we have carried out a sum over polarizations according to Eq.(\ref{eq:polarization_sum}) as in Eq.(\ref{eq:appendix_7}), and  where the sum over ${\bf k}$ in the  large volume $V$ limit  is replaced by
\begin{equation}  
\label{QF_6}
    \sum_{{\bf k}} = \frac{V}{(2\pi)^3} \int d^3k \,\, .
\end{equation}
The  Fourier transform of the transverse current vector in Eq.(\ref{eq:meanfield}) is as above given by
\begin{equation}
   {\bf j}_T({\bf k},t) \equiv  \int d^3x e^{i{\bf k} \cdot {\bf x}} {\bf j}_T({\bf x},t) ~ .
\end{equation}

The time-derivative of Eq.(\ref{eq:meanfield}) can now be written in the form
\begin{gather}  \label{eq:causality15} 
    \frac{\partial \langle {\bf A}_T({\bf x},t) \rangle}{\partial t} = \frac{1}{\epsilon_0}  \int d^3x'  \int_0^t dt' \frac{\partial }{\partial t}G({\bf x}-{\bf x}',t- t')
     \Big ( {\bf j}({\bf x}',t') - {\bf j}_L({\bf x}',t') \Big ) \,\,  ,
\end{gather}
by making use of the  Helmholtz decomposition 
of the current vector ${\bf j}({\bf x},t)$, and where we identified the Green's function $G({\bf x},t)$  
%
\begin{gather} 
    G({\bf x},t) \equiv   \lim_{V\rightarrow \infty} \frac{1}{V}\sum_{{\bf k}} e^{i{\bf k} \cdot {\bf x}} \frac{ \sin( \omega_k t ) }{\omega_k} =\int \frac{d^3k}{(2\pi)^3} e^{i{\bf k} \cdot {\bf x}} \frac{ \sin( \omega_k t ) }{\omega_k}=\nonumber \\ =\frac{1}{4\pi c^2 |{\bf x}|} \bigg ( \delta(t-|{\bf x}|/c) - \delta(t+|{\bf x}|/c) \bigg ) \, .
\label{eq:GreenFunc}
\end{gather}
This  Green's function is a solution to the homogeneous wave-equation 
\begin{equation}
\label{eq:waveoperator}
\frac{\partial^2 {G}({\bf x},t)}{\partial t^2} = c^2 \nabla^2 {G}({\bf x},t)  \,  ,
\end{equation}
such that ${G}({\bf x},0) =0$.
For the second term in Eq.(\ref{eq:causality15}) we need to consider the integral
\begin{equation}
     {\bf I} \equiv    \int d^3x'  \int_0^t dt' \frac{\partial}{\partial t} G({\bf x}-{\bf x}',t- t') \nabla ' \frac{\partial}{\partial t'} \phi ({\bf x}', t')   ~ .
\end{equation}
This is so since the longitudinal vector current ${\bf j}_L({\bf x},t)$ may be written in the form
\begin{equation}
    {\bf j}_L({\bf x},t) = \frac{\partial}{\partial t} \frac{1}{4\pi} \nabla \int d^3x' \frac{ \rho({\bf x}',t) }{|{\bf x}-{\bf x}'|} =  \epsilon_0 \frac{\partial}{\partial t} \nabla \phi({\bf x},t)~ ,
\end{equation}
where we make use of  the Helmholtz decomposition Eq.(\ref{eq:causality13}) and current conservation.
After a partial integration in the time variable $t'$ and by making use of Eq.(\ref{eq:waveoperator}), the integral ${\bf I}$ can therefore be written in the following form
\begin{gather}  
 {\bf I} = \nabla \phi({\bf x},t) + c^2  \int d^3x' \int_0^t dt' \nabla '\cdot \nabla ' G({\bf x}-{\bf x}',t- t')\nabla '\phi({\bf x}',t')  \, , 
\end{gather}
where we have used the fact that $\partial G({\bf x},t)/\partial t =\delta ^3 ({\bf x})$ at $t=0$ as well as the initial condition ${\bf j}_L({\bf x},0)=0$ for all ${\bf x}$.
We now perform two  partial integrations over the spatial variable and by using Eq.(\ref{eq:laplace}) we, finally, see that
\begin{gather}   \label{eq:Integral}
     {\bf I} = \nabla \phi({\bf x},t) - \frac{c^2}{{\epsilon_0}} \int d^3x' \int_0^t dt'  G({\bf x}-{\bf x}',t- t')\nabla '\rho({\bf x}',t')  \, , 
\end{gather}
neglecting  spatial boundary terms and using the initial condition $\rho({\bf x},0)=0$ for all ${\bf x}$.  The first term in Eq.(\ref{eq:Integral})  exactly cancels the instantaneous Coulomb potential contribution in the expectation value of the quantized electric field observable
\begin{equation}
  \langle {\bf E}({\bf x},t)\rangle = - \frac{\partial \langle {\bf A}_T({\bf x},t) \rangle }{\partial t} - \nabla \phi({\bf x},t) ~ .
\end{equation}
For $t > t'$,  we therefore obtain the desired result
\begin{gather}  \nonumber
     \langle {\bf E}({\bf x},t)\rangle = - \frac{\partial}{\partial t} \bigg ( \frac{1}{4\pi \epsilon_0 c^2 } \int d^3x' \frac{ {\bf j} \big ( {\bf x}',t-|{\bf x}-{\bf x}'|/c \big ) }{|{\bf x}-{\bf x}'|} \bigg ) 
     \\
     -  \frac{1}{4\pi \epsilon_0} \int d^3x' \frac{\nabla '\rho \big ( {\bf x}',t-|{\bf x}-{\bf x}'|/c \big )}{|{\bf x}-{\bf x}'|} ~ ,
\label{eq:ave_electric}
\end{gather}
where $\nabla '\rho ({\bf x}',t')$ in Eq.(\ref{eq:ave_electric}) has to be evaluated for a fixed value of $t'= t-|{\bf x}-{\bf x}'|/c$.
 In a similar manner we also see that
\begin{gather} \nonumber 
 \langle {\bf B}({\bf x},t)\rangle   = \nabla \times\langle {\bf A}_T({\bf x},t)\rangle =  \frac{1}{\epsilon_0}\int d^3x' \int_{0}^{t}dt'G({\bf x}-{\bf x}',t- t')\nabla '\hspace{-1.0mm}\times {\bf j}_T({\bf x}',t')   \\
\hspace{-1.5cm}=\frac{\mu _0}{4\pi} \int d^3x'\frac{1} {|{\bf x}-{\bf x}'|} \nabla '\hspace{-1.0mm}\times {\bf j}({\bf x}',t-|{\bf x}-{\bf x}'|/c)\,\, ,
\label{eq:ave_magnetic}
\end{gather}
since $\nabla ' \hspace{-1.0mm}\times{\bf j}_T({\bf x}',t') = \nabla '\hspace{-1.0mm}\times{\bf j}({\bf x}',t')$. In Eq.(\ref{eq:ave_magnetic}), we remark again that $\nabla ' \hspace{-1.0mm}\times{\bf j}({\bf x}',t')$  has to be evaluated for a fixed value of $t'= t-|{\bf x}-{\bf x}'|/c$.
The causal and properly retarded form of the electric and magnetic quantum field expectation values in terms of the physical and local sources given have therefore been obtained (see in this context, e.g., Ref.\cite{Jackson75}, Section 6.5).

The expectation values as given by Eqs.(\ref{eq:ave_electric}) and (\ref{eq:ave_magnetic}) obey Maxwell's equations in terms of the classical charge density $\rho$ and current ${\bf j}$. The quantization procedure above of the electromagnetic field explicitly breaks Lorentz covariance. Since, however, Maxwell's equations transform covariantly under Lorentz transformations we can, nevertheless,  now argue that the special theory of relativity emerges in terms of expectation values of gauge-invariant second-quantized electromagnetic fields.

Maxwell's equations of motion according to Eqs.(\ref{eq:causality1})-(\ref{eq:causality4}) are invariant under the discrete time-reversal  transformation $t \rightarrow t'=-t$  with  ${\bf E}({\bf x},t)  \stackrel{{\cal T}}\longrightarrow   {\bf E}'({\bf x},t)={\bf E}({\bf x},-t)$ and  ${\bf B}({\bf x},t)  \stackrel{{\cal T}}\longrightarrow {\bf B}'({\bf x},t)= - {\bf B}({\bf x},-t)$ provided ${\bf j}({\bf x},t)  \stackrel{{\cal T}}\longrightarrow   {\bf j}'({\bf x},t)=-{\bf j}({\bf x},-t)$ and  $\rho({\bf x},t)  \stackrel{{\cal T}}\longrightarrow \rho'({\bf x},t)= \rho({\bf x},-t)$. 
At the classical level, the corresponding transverse vector potential transforms according to ${\bf A}_T({\bf x},t)  \stackrel{{\cal T}}\longrightarrow   {\bf A}'_T({\bf x},t)=-{\bf A}_T({\bf x},-t)$. 
The anti-unitary time-reversal transformation ${\cal T}$ is implemented on second-quantized fields in the interaction picture  according to the rule (see, e.g., Refs.\cite{Sakurai_1967},\cite{Bjorken_Drell_1965}-\cite{Lopes_1969})
\begin{gather}  \nonumber
\, _I\langle \psi(t)|{\bf A}_T({\bf x},t) |\psi(t) \rangle_I \stackrel{{\cal T}}\longrightarrow \, _I\langle \psi(t)|{\bf A}'_T({\bf x},t) |\psi(t) \rangle_I \\
= \, _I\langle \psi(-t)|{\cal T}{\bf A}_T({\bf x},t) {\cal T}^{-1}|\psi(-t) \rangle_I\,\, .
\label{eq:time_reversal}
\end{gather}
It then  follows that $\, _I\langle \psi(t)|{\bf A}'_T({\bf x},t) |\psi(t) \rangle_I = -\, _I\langle \psi(-t)|{\bf A}_T({\bf x},-t) |\psi(-t) \rangle_I$  if ${\cal T}a_{{\bf k} \lambda}{\cal T}^{-1} = a_{-{\bf k} \lambda}(-1)^{\lambda}$ and provided  the vacuum state $|0\rangle$ is invariant under time-reversal. We therefore find that  $\langle {\bf A}'_T({\bf x},t)\rangle = -\langle {\bf A}_T({\bf x},-t) \rangle$. We therefore obtain  $\langle {\bf E}({\bf x},t)\rangle \stackrel{{\cal T}}\longrightarrow \langle {\bf E}({\bf x},-t)\rangle $ and $\langle {\bf B}({\bf x},t)\rangle \stackrel{{\cal T}}\longrightarrow -\langle {\bf B}({\bf x},-t)\rangle$ as it should. 
Due to the form of the Green's function $G({\bf x},t)$  in Eq.(\ref{eq:GreenFunc}) it  can  be verified   that Eq.(\ref{eq:meanfield})  also leads  to average values  $\langle {\bf E}({\bf x},t)\rangle$ and $\langle {\bf B}({\bf x},t)\rangle$ that transform correctly under time-reversal.  

The arrow of time can therefore, as expected, not be explained by our approach but as soon as the direction of time is defined the observable quantities $\langle {\bf E}({\bf x},t)\rangle$ and  $\langle {\bf B}({\bf x},t)\rangle$ are causal and properly retarded. 
In the presence of external sources we could  have an apparent breakdown of time-reversal invariance unless one also time-reverses  the external sources.
%
%
%
%
%
\vspace{1.2cm}

\bc{
\section{\large Electromagnetic Radiation Processes}
\label{sec:gphoton_emission_proc}
}
\ec
\seqnoll

The rate for  spontaneous emission of a photon from, e.g.,  an excited hydrogen atom can now  be obtained in a straightforward manner  in terms of a slight extension of  the interaction Eq.(\ref{eq:multimode_H}) as to be made  use of in first-order time-dependent perturbation theory.   We then make use of  the long wave-length approximation
\begin{gather} 
\label{eq:qed_identity}
{\bf j}({\bf k},t) \approx \int d^3x {\bf j}({\bf x},t) = \frac{\partial }{\partial t}\int d^3x{\bf x} \rho({\bf x},t) = q \frac{d}{dt}{\bf x}(t)\,\, ,
\end{gather}
taking current conservation Eq.(\ref{eq:conservation}) into account,  where  $\rho({\bf x},t)=q\delta^{(3)}({\bf x}-{\bf x}(t))$ in terms of the position ${\bf x}(t)$ of the charged electron in the interaction picture.  For the spontaneous single photon  transition $\ket{i} \rightarrow \ket{f}$ with  $\ket{i}=\ket{a_i}\otimes\ket{0}$ and $\ket{f} = \ket{a_f}\otimes \ket{{\bf k}\lambda}$, we then arrive at  the standard dipole radiation  first-order matrix element
\begin{gather} 
\label{eq:qed_matrix_element}
\langle i  |H_I(0) |f \rangle = iq \sqrt{\frac{\hbar}{2V\epsilon_0\omega_k}} \omega_{if}{\bm{\epsilon}}({\bf k} ;\lambda)\cdot \langle a_f| {\bf x}|a_i \rangle\,\, ,
\end{gather}
using Eq.(\ref{eq:multimode_H}) with $q=-e$ in the interaction picture. The relevant matrix element  $\langle a_f|{\bf x}(t)|a_i \rangle$ is  then given by $\exp (-i\omega_{if}t)\langle a_f| {\bf x}|a_i \rangle $.  For the atomic transition from $\ket{a_i}=\ket{nlm}= \ket{2pm}$ to the final atomic ground state $\ket{a_f} = \ket{1s}$,  the corresponding rate is then given by
\begin{gather} 
\label{eq:photon_rate}
\Gamma \equiv \frac{2\pi}{\hbar^2}\sum_{{\bf k}\lambda}\delta(\omega_k - \omega_{if})|\langle i|H_I(0) |f \rangle|^2 = \left(\frac{2}{3}\right)^8\alpha ^4\frac{c}{a_B} \,\, ,
\end{gather}
in terms of the fine-structure constant $\alpha \equiv e^2/4\pi \epsilon_0\hbar c$ and the Bohr radius $a_B$. The rate $\Gamma$ is  independent of the quantum number $m$. Stimulated emission gives rise to a multiplicative factor $(1+n_{{\bf k}\lambda})$. Eq.(\ref{eq:photon_rate}) is, of course, a well-known text-book result in agreement with the experimental value (see, e.g., Ref.\cite{Goodman_1960}). The considerations above can be extended to graviton quadrupole radiation processes in an analogous manner \cite{ESR_II}. 

  The power of electromagnetic emission from a classical conserved electric current in, e.g.,  a non-dissipative dielectric medium and  the fameous Vavilov-\v{C}herenkov \cite{Cerenkov_34} radiation can, furthermore, now also be derived in terms of the  quantum-mechanical framework above.  This form of radiation was first explained by Frank and Tamm \cite{Frank_Tamm_37} using the framework of Maxwell's classical theory of electromagnetism. 
The exact classical $\hbar$-independent expression for the power of Vavilov-\v{C}herenkov radiation (see, e.g., Sect. 13.4 in Ref.\cite{Jackson75}), neglecting possible spin effects to be discussed elsewhere \cite{Skagerstam_2017a}, can  now be obtained as follows.
For a particle with electric charge $q$, mass $m$, and an initial velocity ${\bf v}$, moving in  dielectric medium such that $\epsilon_0 \rightarrow \epsilon\epsilon_0$, with $\epsilon >1$, the interaction $H_I(t)$ in Eq.(\ref{eq:multimode_H}) leads to a displacement operator with  $\alpha_{{\bf k}\lambda}(t)$ now replaced by
\begin{gather} 
   \alpha_{{\bf k}\lambda}(t) 
   = \frac{i}{\hbar} \, \sqrt{ \frac{\hbar}{2V\epsilon \epsilon_0 \omega_k} } \, \int_{t_0}^t dt' \, e^{i \omega_k t'}\int {d^3x} {\bf j}({\bf x},t') \cdot {\bm \epsilon}^*({\bf k}; \lambda)e^{-i{\bf k}\cdot{\bf x}} \nonumber \\
 = q\frac{i}{\hbar} \, \sqrt{ \frac{\hbar}{2V\epsilon \epsilon_0 \omega_k} }{\bf v} \cdot {\bm \epsilon}^*({\bf k}; \lambda) \, \int_{t_0}^t dt' \, e^{i (\omega_k  - {\bf k}\cdot{\bf v})t'}~ ,
\label{eq:E_1}
\end{gather}
where the relativistic current ${\bf j}({\bf x},t)$ in an inertial frame is given by 
\begin{gather} 
   {\bf j}({\bf x},t) = q{\bf v}\delta^{(3)}({\bf x}- {\bf v}t)\,\, . 
\label{eq:E_2}
\end{gather}
 The power $P(\omega)d\omega$ of emitted radiation in the range $\omega $ to $\omega +d\omega$ is then obtained by evaluating  the exact expression $d\langle H_0 \rangle(t)/dt$, using Eq.(\ref{eq:E_1}), where
\begin{gather} 
 \frac{d\langle H_0 \rangle(t)}{dt} = \frac{d}{dt}\sum_{{\bf k}\lambda} |\alpha_{{\bf k}\lambda}(t) |^2 \equiv \int_0^{\omega_c} d\omega P(\omega)\,\, , 
\label{eq:E_32}
\end{gather}
 in the large volume $V$ limit, and by considering the large $T\equiv t-t_0$ limit. Here we can, of course, disregard the additive divergent zero-point fluctuations in $\langle H_0 \rangle(t)$.  The large $T$ limit leads to a phase-matching condition $\omega_k = {\bf k}\cdot{\bf v}=vk \cos \theta_C $, using $v\equiv|{\bf v}|$ and $k\equiv|{\bf k}|$,  expressed in terms of the well-known \v{C}herenkov angle $\cos \theta_C \equiv c/nv$, where $\omega_k =ck/n$ with the refractive index $n\equiv \sqrt{\epsilon}$. The $\lambda$-sum over the polarization degrees of freedom in Eq.(\ref{eq:E_32}) leads to 
\begin{gather} 
   \sum_{\lambda = 1,2} |({\bf v}\cdot{\bm \epsilon}({\bf k}; \lambda)|^2 = |{\bf v}|^2(1- \cos^2\theta)\,\, , 
\label{eq:E_4}
\end{gather}
using Eq.(\ref{eq:polarization_sum}), where, in general,  $\cos\theta \equiv {\hat{{\bf v}}}\cdot {\hat{{\bf k}}}$ in terms of the  unit vectors.  
In summing  over the angular distribution of the radiation emitted in Eq.(\ref{eq:E_32}), the large $T$ phase-matching  condition is taken into account. The cut-off angular frequency $\omega_c$ is to be determined in a standard manner taken the $\omega$-dependence of $\epsilon$ into account (see, e.g.,  Ref.\cite{Jackson75}).
We then easily find the well-known $\hbar$-independent power spectrum
\begin{gather} 
   P(\omega) = \frac{e^2}{4\pi \epsilon_0c}\frac{v}{c}\omega\left(1- \cos^2 \theta_C\right)\,\, . 
\label{eq:E_5}
\end{gather}
Alternatively, but in a less rigorous manner, one may consider  $\langle H_0 \rangle(t)/T$ and making use of  Eq.(\ref{eq:E_1}) in the  large $T$-limit, i.e., 
\begin{gather} 
   \alpha_{{\bf k}\lambda}(t) 
   = 2\pi q\frac{i}{\hbar} \, \sqrt{ \frac{\hbar}{2V\epsilon \epsilon_0 \omega_k} } \delta\left( \omega_k - {\bf k}\cdot{\bf v}\right) {{\bf v}\cdot{\bm \epsilon}^*({\bf k}; \lambda)} ~ .
\label{eq:E_3}
\end{gather}
By inspection we then observe that $\alpha_{{\bf k}\lambda}(t)$  in Eq.(\ref{eq:E_3}) exactly corresponds the quantum-mechanical amplitude for the emission of one photon from the source to first-order in time-dependent perturbation theory even though our expression for $ \alpha_{{\bf k}\lambda}(t) $ is exact.

We have therefore derived a power spectrum that exactly corresponds to  the 1937 Frank-Tamm expression \cite{Frank_Tamm_37}  in terms of the \v{C}herenkov angle $\cos \theta_C$  as obtained from the $\delta$-function constraint in Eq.(\ref{eq:E_3}).  In the quantum-mechanical perturbation theory language this constraint corresponds to an energy-conservation $\delta$-function as a well as to conservation of momentum taking the refractive index $n\equiv \sqrt{\epsilon}$ into account. The corresponding energy of the emitted photon is then given by $E_{\gamma} =\hbar\omega$ and   the Minkowski canonical momentum by ${\bf p}_{\gamma} =\hbar {\bf k}$ (see, e.g., Ref.{\cite{Griffiths_2012}),  with $\omega =c|{\bf k}|/n$. The expression for  the \v{C}herenkov angle $\cos \theta_C$ is then modified according to \cite{Ginzburg_1940}
\begin{equation}   \label{eq:rel_cherenkov1}
    \cos \theta_C = \frac{c}{nv} \left ( \, 1 + \hbar \omega_k (n^2-1) \frac{ \sqrt{1-v^2/c^2} }{2mc^2} \, \right ) ~ .
\end{equation}

As was first noted by Ginzburg (\cite{Ginzburg_1940} and references cited therein),  and also presented in various text-books accounts (see, e.g., Refs.\cite{Harris_72, Marcuse_80}), first-order perturbation theory in quantum mechanics  actually leads to the same exact power spectrum for Vavilov-\v{C}herenkov radiation. The explanation of this curious circumstance can be traced back to the fact that all higher order corrections are taken into account by the presence of the phase $\phi(t)$ in Eq.(\ref{eq:timeevolution}).
%
%
%
%
\vspace{0.2cm}
\bc{
\section{Quantum Uncertainty}
\label{sec:fluctuations}
}\ec
\seqnoll

The displacement of quantum states as induced by $D({\bm \alpha})$, as defined in Eq.(\ref{eq:def_of_D}), acting on an arbitrary pure initial state  again leads to Maxwell's equations for  the expectation value of the  quantum field changing,  at most, the homogeneous solution of the expectation value of the  wave-equation (\ref{eq:average_motion}). The corresponding quantum uncertainty of ${\bf E}({\bf x},t)$, however, depends on the choice of the initial state along the same reasoning as in the single-mode case in Section \ref{sec:one_mode_only}. An essential and additional ingredient with regard to the approach to the  classical limit is to consider the variance of, e.g.,  the second-quantized ${\bf E}({\bf x}, t)$-field suitably defined. We consider the scalar quantity
\begin{eqnarray}
\label{QF_0}
(\Delta {\bf E}({\bf x}, t))^2 \equiv  \langle{\bf E}^2({\bf x}, t) \rangle - \langle
{\bf E}({\bf x}, t)\rangle ^2   \, \, .
\end{eqnarray}
We observe that the uncertainty in Eq.(\ref{QF_0}) in general does not depend on the complex parameters ${\bm \alpha}$ when evaluated for  the displaced state $D({\bm \alpha})\ket{\psi (0)}$ and is therefore determined  by the uncertainty as determined by the initial state $\ket{\psi (0)}$.

In order to be specific, we will evaluate the uncertainty $\Delta {\bf E}({\bf x}, t)$ for a displaced  Fock state with $\ket{\psi (0)}= |..., n_{{\bf k} \lambda}, ....,n_{{\bf k}' \lambda'} , .. \rangle$. We then obtain
\begin{eqnarray}
\label{QF_1}
(\Delta {\bf E}({\bf x}, t))^2  
= \frac{\hbar}{V \varepsilon_0}\sum_{{\bf k} \lambda} \omega_{k}(n_{{\bf k} \lambda}+ \frac{1}{2}) \, \, .
\end{eqnarray}
 Physical requirements now  demand that the uncertainty $\Delta {\bf E}({\bf x}, t)$ must be smaller than expectation values of the components of the second-quantized electromagnetic field ${{\bf E}({\bf x}, t)}$. If the sum in Eq.(\ref{QF_1}) had been convergent, the variance would  have vanished in the naive limit $\hbar\to 0$.  Since the natural constant $\hbar$ is non-zero, the sum in 
Eq.(\ref{QF_1}) is, however, divergent.

Even though the expectation value of the quantum field at a space-time point $({\bf x}, t)$ in our case is well-defined, the corresponding uncertainty is therefore actually divergent. This means that the observable value of the quantum field in a space-time point $({\bf x}, t)$ is physically ill-defined. In the early days of quantum field theory, this fact was actually noticed already in 1933 by Bohr and Rosenfeld \cite{Bohr_1950} and later proved in a rigorous manner by Wightman \cite{Wightman_1964}.  Bohr and Rosenfeld also provided  a solution of this apparent physical contradiction.  The basic idea is to introduce quantum field observables
averaged over some finite space-time volume. Bohr and Rosenfeld made use of a cube centered at the space-point ${\bf x}$ at a fixed time $t$ which, however, makes some of the expressions obtained rather complicated. We will follow another approach which makes the expressions more tractable (see, e.g.,  problem $2.3$ in Ref.\cite{Sakurai_1967}), i.e.,  we consider
\begin{equation}
\label{QF_2}
 {\bf E}_{\sigma}({\bf x}, t) \equiv \int_{-\infty}^{\infty}dt'\int_V d^3x' f_{s}({\bf x}-{\bf x}')f_{t}(t-t')   {\bf E}({\bf x}', t')\, ,
\end{equation}
where 
\begin{equation}
\label{QF_3}
f_{s}({\bf x}) = \frac{1}{(2\pi \sigma_s^2)^{3/2}} \exp (-\frac{{\bf x}^2}{2\sigma_s^2}) \, ,
\end{equation}
and
\begin{equation}
\label{QF_33}
f_{t}(t) = \frac{1}{(2\pi \sigma_t^2)^{1/2}} \exp (-\frac{t^2}{2\sigma_t^2}) \, .
\end{equation}
The parameter $\sigma_{s}$ gives a characteristic scale for the space-volume around the point ${\bf x}$ where we perform the space average. Correspondingly, the parameter $\sigma_{t}$ gives a characteristic  time-scale for the time average procedure. 

The linear classical Maxwell's  equations can then again be obtained as in the previous sections in terms of the quantum-mechanical average of fields like ${\bf E}_\sigma({\bf x}, t)$ provided that the classical sources are space and/or time averaged in the same manner. It now follows that
\begin{gather}
\label{QF_4}
  {\bf E}_\sigma({\bf x}, t)= \sum_{{\bf k} \lambda}i \sqrt{\frac{\hbar \omega_k}{2V\varepsilon_0}}\exp(-\frac{\sigma_s^2{\bf k}^2}{2}-\frac{\sigma_t^2{\omega_k}^2}{2} )  \nonumber \\ 
 \times \bigg (  a_{{\bf k} \lambda} {\bm \epsilon}({\bf k};\lambda) e^{i( {\bf k} \cdot {\bf x}- \omega_k t)} - a^*_{{\bf k} \lambda} {\bm\epsilon}^*({\bf k};\lambda) e^{-i ({\bf k} \cdot 
{\bf x}-\omega_k t)}  \bigg )\,\, .
\end{gather}
The variance $(\Delta {\bf E}_{\sigma}({\bf x}, t))^2$ of the space and time averaged electric quantum field  ${\bf E}_\sigma({\bf x}, t)$, in a sufficiently large quantization volume $V$,  will then be finite and corresponds to an energy 
\begin{eqnarray}
\label{QF_5}
 \varepsilon_0(\Delta {\bf E}_\sigma({\bf x}, t))^2 \sigma ^3 \geq
 E_{\sigma} \equiv \frac{1}{(2\pi)^3} \frac{2\pi \hbar c}{\sigma} \, ,
\end{eqnarray}
localized in a volume $V_{\sigma} \equiv \sigma^3 \ll V$, 
where
\begin{eqnarray}
\label{QF_55}
\sigma^2 \equiv \sigma_s^2+c^2\sigma_t^2 \,\, .
\end{eqnarray}
It is now clear that $E_{\sigma}$ will be finite in the cases $\sigma_{t}=0$, $\sigma_{\text{s}} \neq 0$ as well as $\sigma_{s}=0$, $\sigma_{t} \neq 0$.

  Physically, the expression in Eq.(\ref{QF_5}) corresponds, apart from an irrelevant numerical factor, to the energy of a photon with a wave-length $\lambda \simeq \sigma$ and, hence,   a wave-number $k\simeq 2\pi /\sigma$ and therefore to an energy $E_\sigma \simeq \hbar c k \simeq 2\pi\hbar c /\sigma$,  in a typical localization volume $V_{\sigma}$.  It is now clear that  $E_\sigma$ will tend to infinity as $\sigma \rightarrow 0$, i.e., we would then obtain an arbitrarily large energy and/or energy density if we try to localize the quantum field in the sense above  in  an arbitrary small $V_{\sigma}$. A macroscopic field, however,  corresponds to a localization volume much larger than $V_{\sigma}$, and therefore these quantum uncertainties can be disregarded in the classical regime.  

This latter feature can be illustrated by evaluating $(\Delta {\bf E}_{\sigma}({\bf x}, t))^2$ for a thermal Planck distribution of $n_{{\bf k} \lambda}$ at a temperature $T$ with a typical coherence length scale $\sigma_T \equiv \hbar c /k_BT$. For localization scales $\sigma \ll \sigma_T$, i.e., at sufficiently small temperatures, one then finds that  $\varepsilon_0(\Delta {\bf E}_\sigma({\bf x}, t))^2 \sigma ^3= E_{\sigma}(1+ 4\pi^4(\sigma/\sigma_T)^4/15)$. The thermal induced  uncertainty can therefore be neglected in comparison with $E_{\sigma}$ for large thermal coherence lengths $\sigma_T$ as compared to $\sigma$.  If, on the other hand, $\sigma_T \leq \sigma$, i.e., at sufficiently high temperatures,  it follows that $\varepsilon_0(\Delta {\bf E}_\sigma({\bf x}, t))^2 \sigma ^3 =k_BT(1+(\sigma_T/\sigma)^2/8)/4\pi^{3/2}$ and, as expected, the thermal uncertainty will then be dominating  at sufficiently high temperatures. 

As was predicted a long time ago  for single-mode quantum fields \cite{{Walls_Zoller_1981}},  it is possible to reduce the uncertainty $(\Delta {\bf E}({\bf x}, t))^2$ below the vacuum value by making use of  initial squeezed quantum states $|\psi(0) \rangle$. This feature has recently been confirmed experimentally (Ref.\cite{{Schule_2015}} references cited therein). For multi-mode considerations, relevant for the framework of the present work, this may also be possible for $(\Delta {\bf E}_{\sigma}({\bf x}, t))^2$ but this will not be a topic in the present paper.
%
%
\vspace{0.2cm}

\bc{
\section{\large Final Remarks}
\label{sec:final_remarks}
}
\ec
\seqnoll

We have seen that a quantum-mechanical  framework offers a platform to study causality and retardation issues in the classical theory of Maxwell. As has been shown elsewhere,  this framework can rather easily be extended to  a derivation of the weak-field limit of Einstein's general theory of relativity \cite{ESR_II}. From second-quantization of the physical degrees of freedom under the condition of  current conservation the well established classical theory for electromagnetism naturally emerges. The overwhelming experimental support for Maxwell's classical theory does not necessarily imply the existence of photons and doubts on the existence of such quantum states are sometimes put forward (see, e.g., Ref.\cite{Lamb_95}). However, the quantum-mechanical derivation of the classical theory necessarily  implies the existence of single particle quantum states corresponding to a photon. 

We have also observed that various radiation processes  including the  classical, i.e., $\hbar$-independent,   Vavilov-\v{C}herenkov radiation can be obtained  in a straightforward manner. 
It may come as  a surprise that a first-order quantum-mechanical perturbation theory calculation can give an exact $\hbar$-independent answer. This, as it seems,  remarkable fact is explained by  the factorization of the time-evolution operator in terms of a displacement operator for  quantum states in the interaction picture according to Eq.(\ref{eq:timeevolution}) making use of Eq.(\ref{eq:multimode_H}). The phase $\phi(t)$ then contains the non-perturbative effects of all higher-order corrections to the first-order result. 

As a matter of fact, similar features are known to occur also in some other situations. As is well-known, the famous differential cross-section for  Rutherford scattering can be obtained exactly in terms of the first-order Born approximation. All higher order corrections will then contribute with an overall phase for probability amplitudes which follows from the exact solution (see, e.g., the excellent discussion in Ref.\cite{Gottfrid_66}). The classical Thomson cross-section for low-energy light scattering on a charged particle is also exactly obtained from a Born approximation due to the existence of an exact low-energy theorem in quantum electrodynamics (see, e.g., the discussions in  Refs.\cite{Itzykson_1972, Itzykson_1980, Thirring_1980}).
\vspace{1cm}
%
%
%
\renewcommand{\thesection}{A}
\renewcommand{\section}[1]{\refstepcounter{sectionc}\vspace{0.0cm}
\setcounter{subsectionc}{0}\setcounter{subsubsectionc}{0}\noindent 
	{\bf\thesectionc. #1}}
\renewcommand{\theequation}{\thesection.\arabic{equation}}
\setcounter{section}{1}
\begin{center}
{\Large {\sc Appendix }}
\end{center}
\vspace{0.5cm}
%
%
%
%
\begin{center}
{ {\Large \bf \thesection. 
 { Expectation Value of the Quantum Field ${\bf A}_T({\bf x},t)$}}}
\end{center}
\vspace{0.5cm}

 Apart from a phase-factor, the time-evolution in the interaction picture is controlled by the operator
\begin{equation}
   \exp \left ( - \frac{i}{\hbar} \int_{0}^t dt' \, H_I(t') \right )  = \prod_{{\bf k}\lambda} \, \exp \Big ( \alpha_{{\bf k}\lambda}(t) a^*_{{\bf k}\lambda} - {\alpha}^*_{{\bf k}\lambda}(t) a_{{\bf k}\lambda} \Big ) ~ ,
\end{equation}
where $\alpha_{{\bf k}\lambda}(t)$ by is given by Eq.(\ref{eq:alpha}) in the main text.
Since expectation values are independent of the picture used, i.e.,
\begin{equation}
    _S\langle \psi(t)| {\cal O} |\psi(t) \rangle_S \, = \, _I\langle \psi(t)| {\cal O}_I(t) |\psi(t) \rangle_I \equiv  \langle {\cal O}_I(t) \rangle  ~ ,
\end{equation}
we see that
\begin{gather}  
   \langle {\bf A}_T({\bf x},t) \rangle  = \sum_{{\bf k}\lambda} \,  \sqrt{ \frac{\hbar}{2V \omega_k \epsilon_0} }  \, \bigg ( \, {\bm \epsilon}({\bf k};\lambda)\alpha_{{\bf k} \lambda}(t) \, e^{i {\bf k} \cdot {\bf x} - i \omega_k t} + {\bm \epsilon}^*({\bf k};\lambda){\alpha}^*_{{\bf k} \lambda}(t) \, e^{-i {\bf k} \cdot {\bf x} + i \omega_k t} \, \bigg )\,\, .
\end{gather}
If we, in particular, consider initial states $|\psi(0) \rangle$ such that $\langle \psi(0)|{a}_{{\bf k} \lambda}|\psi(0) \rangle = 0$ for all ${\bf k} \lambda$, like quantum states with a fixed number of photons, 
we find that
\begin{gather}  \nonumber
    \langle {\bf A}_T({\bf x},t) \rangle  = \sum_{{\bf k}\lambda} \,  \frac{i}{2V \omega_k \epsilon_0} \, \bigg ( \, {\bm \epsilon}({\bf k}; \lambda)  \int_0^t dt' \, e^{i \omega_k(t'-t) + i{\bf k} \cdot {\bf x}} \,  {\bm \epsilon}^*({\bf k}; \lambda) \cdot {\bf j^*}({\bf k},t') 
    \\ \nonumber
    - {\bm \epsilon}^*({\bf k}; \lambda)\int_0^t dt' \, e^{-i \omega_k(t'-t) - i{\bf k} \cdot {\bf x}} \,  {\bm \epsilon}({\bf k}; \lambda)  \cdot {\bf j}({\bf k},t') \, \bigg ) 
    \\ \label{eq:causality19}
    = - \sum_{{\bf k}\lambda} \, \frac{1}{V \epsilon_0} \, e^{ i {\bf k} \cdot {\bf x} } \, \int_0^t dt' \; \frac{ \sin \Big ( \omega_k (t'-t) \Big ) }{\omega_k} \; {\bm \epsilon}({\bf k}; \lambda) \, \Big ( \, {\bm \epsilon}^*({\bf k};\lambda) \cdot {\bf j}^*({\bf k},t') \, \Big ) ~ ,
\end{gather}
after a change ${\bf k} \rightarrow - {\bf k}$ in the last term above, using ${\bf j}(-{\bf k},t)={\bf j}^*({\bf k},t)$ as well as Eq.(\ref{eq:alpha}). The second time-derivative of this expression will then contain the following factor:
\begin{gather} \nonumber
   \sum_{{\bf k}\lambda} \, \frac{1}{V \epsilon_0} \, e^{ i {\bf k} \cdot {\bf x} } \, {\bm \epsilon}({\bf k}; \lambda) \, \Big ( \, {\bm \epsilon}^*({\bf k}; \lambda) \cdot {\bf j}^*({\bf k},t) \, \Big )
   \\  \label{eq:appendix_7}
   =  \sum_{{\bf k}} \, \frac{1}{V \epsilon_0} \, e^{ i {\bf k} \cdot {\bf x} } \, \bigg ( \, {\bf j}^*({\bf k},t) - \hat{{\bf k}} \Big ( \hat{{\bf k}} \cdot {\bf j}^*({\bf k},t) \Big ) \, \bigg ) ~ ,          
\end{gather}
where ${\bf j}^*_{T}({\bf k},t)={\bf j}_{T}(-{\bf k},t)\equiv {\bf j}(-{\bf k},t) - \hat{{\bf k}} \big ( \, \hat{{\bf k}} \cdot {\bf j}(-{\bf k},t) \, \big )$ corresponds to the Fourier-components of a transverse current ${\bf j}_T({\bf x},t)$ such that $\nabla \cdot {\bf j}_T({\bf x},t) = 0$, and where use have been made of Eq.(\ref{eq:polarization_sum}).
The transverse term obtained using Eq.(\ref{eq:appendix_7}) can therefore be written in the form:
\begin{gather}  \nonumber
     \frac{1}{V \epsilon_0} \, \sum_{{\bf k}} \, e^{ i {\bf k} \cdot {\bf x} } \, {\bf j}_{T}(-{\bf k},t)
     \\
     =  \frac{1}{V \epsilon_0} \, \sum_{{\bf k}} \, e^{ i {\bf k} \cdot {\bf x} } \, \int d^3x' \, e^{- i {\bf k} \cdot {\bf x}'} \, {\bf j}_T({\bf x}',t) = \frac{1}{\epsilon_0} \; {\bf j}_T({\bf x},t) ~ ,
\end{gather}
where we make use of the fact that 
\begin{equation}
     \frac{1}{V} \, \sum_{ {\bf k} } \, e^{i {\bf k} \cdot ({\bf x} - {\bf x}') } = \delta^{(3)}({\bf x} - {\bf x}') ~ .
\end{equation}
We have therefore reproduced the source-term in the wave-equation Eq.(\ref{eq:average_motion}) in the main text
and all the terms we have left out above in the evaluation of $\partial^2 \langle {\bf A}_T({\bf x},t) \rangle/\partial t^2$ satisfy the homogeneous wave-equation.
%
%
%
%
%
%
%
%
%
%
%
\vspace{0.8cm}
%
%
%
%
%
%
%
\begin{center}
   {\bf \large ACKNOWLEDGMENT}
\end{center}
The research by B.-S. S.  was supported in part by Molde University College and NTNU. K.-E. E. gratefully acknowledges hospitality at Chalmers University of Technology. The research of P. K. R. was supported  by  Molde University College. The authors are grateful for the hospitality provided  for at Department of Space, Earth and Environment,
Chalmers University of Technology, Sweden,  during the finalization of the present work.
%
%
%
%
%
%
%
%
%
%
\begin{center}
   {\bf \large REFERENCES}
\end{center}
%
 
%
%
%
%
%
%
%
\end{document}